\documentclass[useAMS,usenatbib]{mn2e}
\usepackage{graphicx}
\usepackage{amssymb,amsmath}
\usepackage{pdflscape}
\usepackage{subfigure}
\usepackage{color}
\usepackage{rotating}
\usepackage{epstopdf}
\usepackage{lscape}
\usepackage{longtable}

\newcommand{\UV}{\emph{uv}\ }

\def\UW{$^{1}$}
\def\Haystack{$^{2}$}
\def\MIT{$^{3}$}
\def\Curtin{$^{4}$}
\def\USwinburne{$^{5}$}
\def\CfA{$^{6}$}
\def\ASU{$^{7}$}
\def\CSIRO{$^{8}$}
\def\RRI{$^{9}$}
\def\USydney{$^{10}$}
\def\CAASTRO{$^{11}$}
\def\UWisc{$^{12}$}
\def\UMelbourne{$^{13}$}
\def\UTasmania{$^{14}$}
\def\ANU{$^{15}$}
\def\PerthUWA{$^{16}$}

\begin{document}

\title[A new optimization technique for interferometric arrays and the final MWA antenna layout]{A new layout optimization technique for interferometric arrays and the final MWA antenna layout}

%Author list
\author[A.~P.~Beardsley, et al.]{
A.~P.~Beardsley\UW, 
B.~J.~Hazelton\UW, 
M.~F.~Morales\UW\thanks{Email: mmorales@phys.washington.edu}, 
R.~C.~Cappallo\Haystack, 
R.~Goeke\MIT,
\newauthor 
D.~Emrich\Curtin, 
C.~J.~Lonsdale\Haystack, 
W.~Arcus\Curtin, 
D.~Barnes\USwinburne, 
G.~Bernardi\CfA, 
J.~D.~Bowman\ASU, 
\newauthor
J.~D.~Bunton\CSIRO, 
B.~E.~Corey\Haystack, 
A.~Deshpande\RRI, 
L.~deSouza\CSIRO$^,$\USydney,
B.~M.~Gaensler\USydney$^,$\CAASTRO, 
\newauthor
L.~J.~Greenhill\CfA,
D.~Herne\Curtin, 
J.~N.~Hewitt\MIT, 
D.~L.~Kaplan\UWisc, 
J.~C.~Kasper\CfA, 
\newauthor
B.~B.~Kincaid\Haystack, 
R.~Koeing\CSIRO, 
E.~Kratzenberg\Haystack, 
M.~J.~Lynch\Curtin, 
S.~R.~McWhirter\Haystack,
\newauthor
D.~A.~Mitchell\CfA$^,$\CAASTRO, 
E.~Morgan\MIT, 
D.~Oberoi\Haystack, 
S.~M.~Ord\CfA, 
J.~Pathikulangara\CSIRO, 
\newauthor
T.~Prabu\RRI, 
R.~A.~Remillard\MIT, 
A.~E.~E.~Rogers\Haystack, 
A.~Roshi\RRI, 
J.~E.~Salah\Haystack, 
\newauthor
R.~J.~Sault\UMelbourne, 
N.~Udaya~Shankar\RRI, 
K.~S.~Srivani\RRI, 
J.~Stevens\CSIRO$^,$\UTasmania, 
R.~Subrahmanyan\RRI, 
\newauthor
S.~J.~Tingay\Curtin$^,$\CAASTRO, 
R.~B.~Wayth\Curtin$^,$\CfA, 
M.~Waterson\Curtin$^,$\ANU,
R.~L.~Webster\CAASTRO$^,$\UMelbourne, 
\newauthor
A.~R.~Whitney\Haystack, 
A.~Williams\PerthUWA, 
C.~L.~Williams\MIT, 
J.~S.~B.~Wyithe\CAASTRO$^,$\UMelbourne
\\
$^{1}$University of Washington, Seattle, USA\\
$^{2}$MIT Haystack Observatory, Westford, USA\\
$^{3}$MIT Kavli Institute for Astrophysics and Space Research, Cambridge, USA\\
$^{4}$Curtin University, Perth, Australia\\
$^{5}$Swinburne University of Technology, Melbourne, Australia\\
$^{6}$Harvard-Smithsonian Center for Astrophysics, Cambridge, USA\\
$^{7}$Arizona State University
$^{8}$CSIRO Astronomy and Space Science, Australia\\
$^{9}$Raman Research Institute, Bangalore, India\\
$^{10}$University of Sydney, Sydney, Australia\\
$^{11}$ARC Centre of Excellence for All-sky Astrophysics (CAASTRO)\\
$^{12}$University of Wisconsin--Milwaukee, Milwaukee, USA\\
$^{13}$The University of Melbourne, Melbourne, Australia\\
$^{14}$University of Tasmania, Hobart, Australia\\
$^{15}$The Australian National University, Canberra, Australia\\
$^{16}$Perth Observatory, Perth, Australia, and the University of Western Australia
}

\label{firstpage}
\pagerange{\pageref{firstpage}--\pageref{lastpage}}\pubyear{2011}

\maketitle

\begin{abstract}

Antenna layout is an important design consideration for radio interferometers because it determines the quality of the snapshot point spread function (PSF, or array beam). This is particularly true for experiments targeting the 21~cm Epoch of Reionization signal as the quality of the foreground subtraction depends directly on the spatial dynamic range and thus the smoothness of the baseline distribution. Nearly all sites have constraints on where antennas can be placed---even at the remote Australian location of the MWA (Murchison Widefield Array) there are rock outcrops, flood zones, heritages areas, emergency runways and trees. These exclusion areas can introduce spatial structure into the baseline distribution that enhance the PSF sidelobes and reduce the angular dynamic range. In this paper we present a new method of constrained antenna placement that reduces the spatial structure in the baseline distribution. This method not only outperforms random placement algorithms that avoid exclusion zones, but surprisingly outperforms random placement algorithms without constraints to provide what we believe are the smoothest constrained baseline distributions developed to date. We use our new algorithm to determine the final antenna placement for the MWA, and present the planned antenna locations, baseline distribution, and snapshot PSF for the observatory.
\end{abstract}


\clearpage

\section{Introduction}
Antenna placement is a critical design criterion for any interferometric array as it determines the baseline distribution and thus the angular dynamic range of the point spread function of the observatory. Nearly all observatory sites have areas where antennas cannot be placed. Buildings, roads, runways, power and data access, land use and ownership issues, endangered flora and fauna, flood zones, elevation, and ground stability are but a few of the common issues that constrain the placement of antennas. Even in remote desert locations a flat and barren terrain can quickly become dotted with exclusion zones where antennas cannot be placed.

This is of particular concern for 21~cm Cosmology telescopes targeting the Epoch of Reionization (EoR) and Baryon Acoustic Oscillation (BAO) dark energy measurements, as the quality of the monochromatic PSF is directly related to the ability to subtract foreground contamination \citep[][Vedantham, Shankar, Subrahmanyan in review]{Morales:2006p147,Datta:2011p4788,Liu:2011p4789,Bernardi:2011}. Antenna exclusion zones can introduce asymmetries in the baseline distribution which limit the angular dynamic range and thus achievable level of foreground subtraction \citep[see][for a recent review of foreground subtraction for 21~cm Cosmology]{Morales:2010p4786}.

%There is a long history of array configuration studies, including optimization of arrays with cost constraints \cite[e.g.\ ][]{Cohanim:2010p4842}, optimization to match a particular baseline distribution without ground constraints \citep[][Lal et\ al.\ SKA Memo 107]{Boone:2001p4841}, or optimization to reduce the peak sidelobe levels \citep[][Kogan \& Cohen 2005, LWA Memo 21]{Kogan:2000p4790}. Our particular concern is situations in which some areas cannot be used (exclusion zones), a particular radial baseline distribution must be met, and a very high angular dynamic range must be achieved.

There is a long history of array configuration studies, including optimization of arrays with cost constraints \cite[e.g.\ ][]{Cohanim:2010p4842}, simulated annealing for small $N$ arrays \citep[]{Cornwell:1988p1165}, optimization to reduce the peak sidelobe levels \citep[][Kogan \& Cohen 2005, LWA Memo 21]{Kogan:2000p4790}, or optimization to match a particular baseline distribution with and without ground constraints \citep[][Lal et\ al.\ SKA Memo 107]{Boone:2001p4841}.  Our particular concern is situations in which some areas cannot be used (exclusion zones), a particular radial baseline distribution must be met, and a very high angular dynamic range must be achieved.  While our problem is similar to that of \cite{Boone:2001p4841}, we find that the figure of merit used in that work does not sufficiently capture large scale structure in the baseline distribution.  We develop an alternative figure of merit, which naturally leads to a new optimization method.

In \S\ref{TechniqueSec} we explore the effect of exclusion zones on the baseline distribution, develop a new spatially sensitive figure of merit, and present our new optimization method. We then apply our method to placing the MWA tiles in \S\ref{sec:final_layout}, and present the final MWA antenna configuration. The locations of all 512 tiles are provided in the electronic supplement.

\section{Array Layout Comparisons and a new technique}\label{TechniqueSec}
Proposed and future large $N$ radio arrays will face the challenge of placing hundreds to thousands of antennas to optimize scientific goals, while obeying numerous constraints. While most physical constraints exist on the antenna locations (areas of exclusion on the ground), science capabilities are optimized in the \UV plane for an interferometric measurement, and hence these arrays should match the ideal \emph{baseline} distribution as closely as possible.  This makes the problem very non-linear because any one antenna placement affects $N-1$ baselines, and it is not immediately obvious how a constraint such as an exclusion area will affect the baseline distribution.

In our analysis, we explored three array layout methods.  The first method is random and with no exclusion areas (``random unmasked''), in which antennas are placed randomly with a weighted radial distribution. Algorithmically, for each antenna a radius is first drawn from a distribution that matches the desired radial density profile, then azimuthal angles are chosen at random until one is found that does not overlap with previously placed antennas. The second array generation method is also random but incorporates exclusion areas (``random masked'').  This method is identical to the random unmasked method, with the addition of avoiding exclusion areas by the use of a mask that is checked in the same step as checking for overlap with previous tiles. The third and final method is the algorithm that we developed (``active method''), that actively minimizes spatial structure in the baseline distribution and is detailed later in this section.  Our analysis assumes the scientifically desired \UV or antenna distribution is known.  We use the specifications for the MWA telescope (which are described in \S \ref{sec:final_layout}), but our methods are generalizable to any large $N$ array. 

\begin{figure*}
\begin{center}
\subfigure{\includegraphics{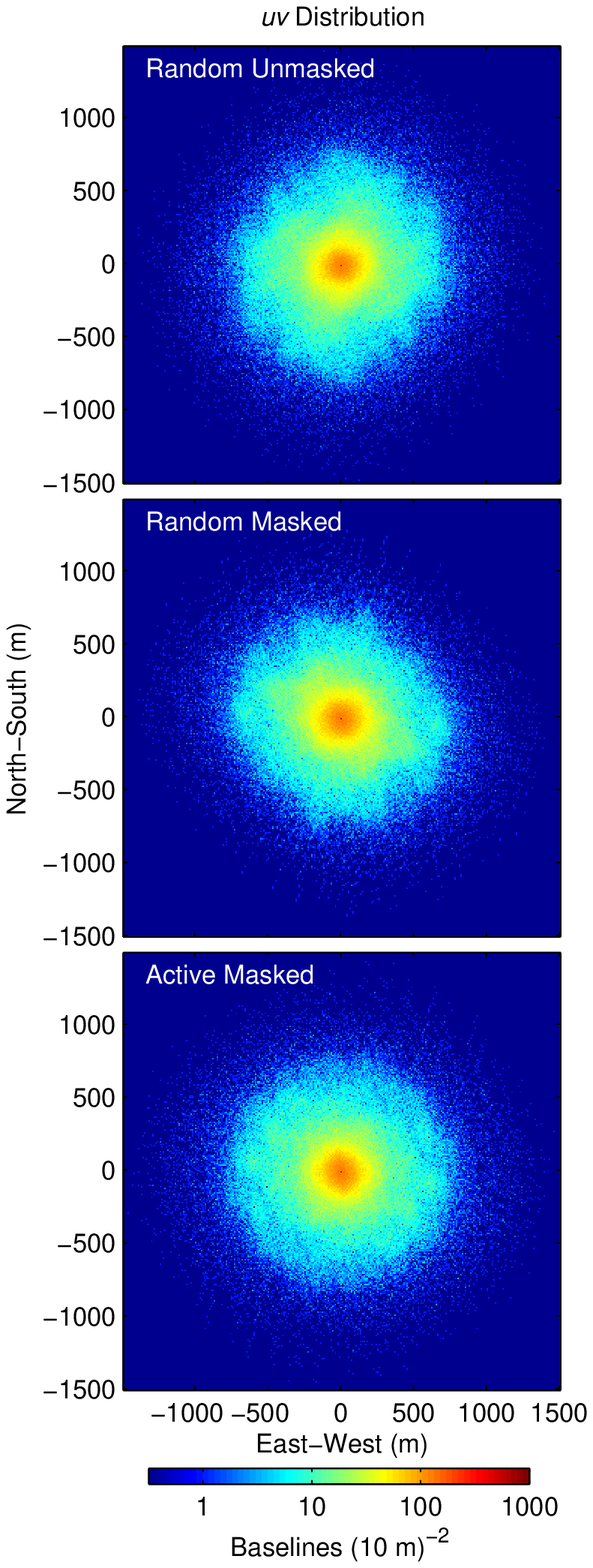}}
\subfigure{\includegraphics{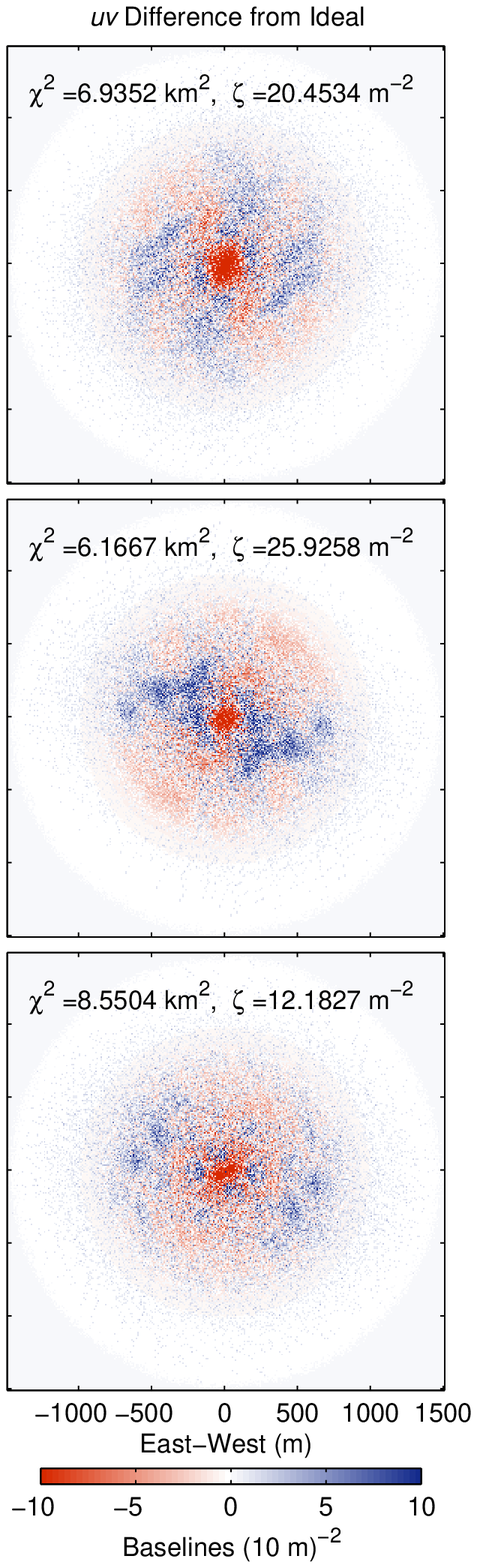}}
\end{center}
\caption{Example \UV distributions for array layouts generated by (top to bottom) the random unmasked, random masked, and active masked methods.  The left panes show the snapshot single frequency baseline distributions on a log scale, while the right panes show the difference of this distribution from the smooth analytic ideal. The small scale fuzzy noise is equally present in all array realizations and is due to the finite number of antennas. However, the large scale  structure varies greatly from array to array. The exclusion areas have introduced significant asymmetries in the baseline distribution of the random masked method (middle row). The active masked method (bottom row) is able to highly suppress this structure, even beyond the level of the \emph{unconstrained} random method (top row).  Furthermore, we see the figure of merit, $\zeta$, accurately reflects the amount of azimuthal structure in the distributions (see Figure \ref{fig:hists}).}
\label{fig:examples}
\end{figure*}

Figure \ref{fig:examples} shows three examples of baseline distributions generated by the random unmasked, random masked, and active masked methods respectively.  The left pane shows the baseline distribution on a logarithmic scale, while the right pane shows the difference from the ideal smooth analytic function to accentuate undesired structures in the \UV distributions.  These three examples are representative of the over four thousand array layouts we have hand graded to arrive at our conclusions. 

% Figure with all the examples:

\begin{figure*}
\begin{center}
\subfigure{
\includegraphics{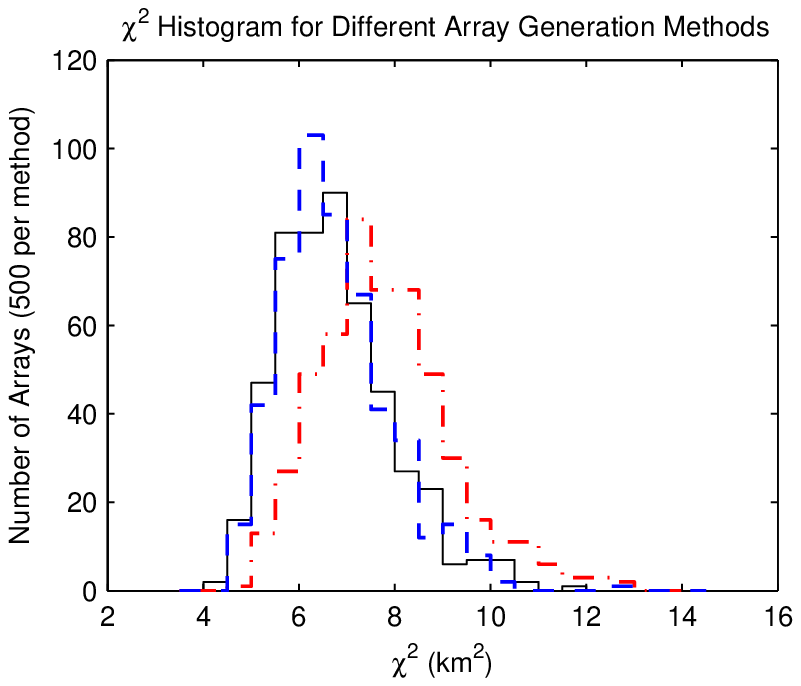}
\label{fig:chi_squared_hist}}
\subfigure{
\includegraphics{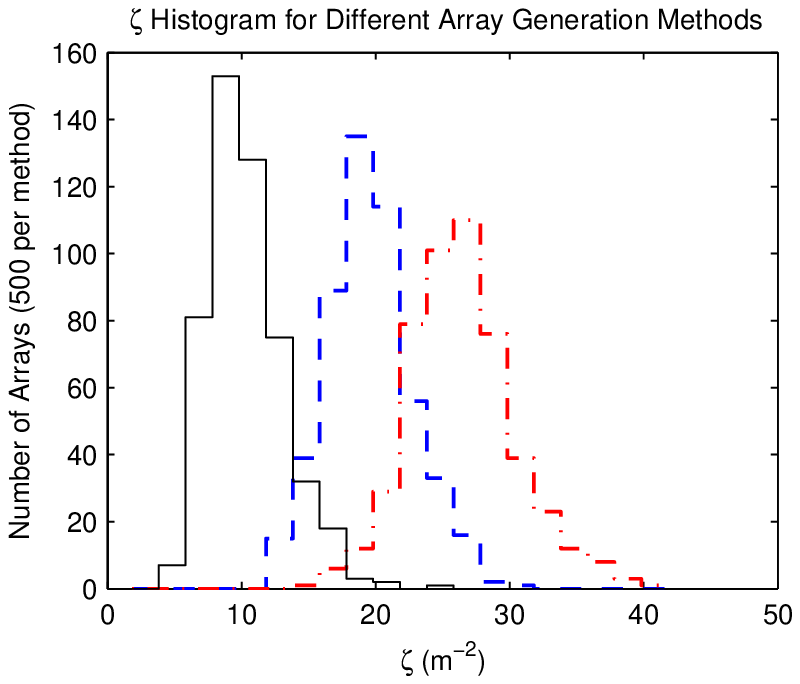}
\label{fig:power_hist}}
\end{center}
\caption{$\chi^{2}$ and $\zeta$ histograms for random unmasked, random masked, and active masked methods as denoted by the thick dashed blue, thick dot-dashed red, and thin solid black lines respectively. While the $\chi^2$ values do not distinguish the quality of the different realizations (a conclusion firmly supported by our hand grading), the $\zeta$ values strongly separate the realizations based on asymmetry. }
\label{fig:hists}
\end{figure*}

All of these images have fuzzy small scale noise due to the finite number of antennas. However, there is another more insidious artifact present in the masked baseline distribution (middle row of Figure \ref{fig:examples})---large scale structure imprinted by the antenna exclusion areas. In all of the masked random array realizations there are significant regions of over and under densities in the baseline distribution which translate directly into unwanted PSF features. 

To understand the effect of baseline over and under densities, consider a nearly perfect \UV distribution with a small region of excess baselines. This region of \UV over density can be viewed as a `wave packet' of baselines at similar spatial frequencies. In the wave packet picture there is a fundamental corrugation in the PSF given by the location of the center of the excess region. However, the nearby modes in the packet beat in and out of phase with the fundamental corrugation. When the wave numbers are all in phase the amplitude is very high---the number of excess baselines in the region---but they quickly dephase only to rephase again some distance further along in the PSF. The undesirable `features' seen in most PSFs are the periodic signature of a wavepacket beating across the PSF. Over (or under) dense regions that cover a large portion of the \UV plane will quickly damp down (wide bandwidth), though they often have a lot of power due to the large number of baselines involved, and correspond to large close-in sidelobes. Smaller features in the \UV plane damp more slowly and repeat many times across the PSF leading to the small far sidelobes. A smooth \UV distribution necessarily leads to a smooth PSF, and the PSF slidelobe structure is dominated not by the unavoidable fuzzy noise but instead the larger regions of over and under density in the \UV plane.

Our first approach to quantify the deviations from the desired \UV distributions was to consider $\chi^2$.  This was calculated by gridding the \UV distribution and integrating the square of the difference from the ideal, weighted by the variance in each pixel from 500 random unmasked realizations.  However, $\chi^2$ is not a spatially aware function---any deviation from the ideal is weighted the same regardless of where in the \UV plane the deviation occurs. Because of this lack of spatial information, $\chi^2$ does not capture the large scale structure that is important to choosing an array. The examples in Figure \ref{fig:examples} vary quite a bit in quality, however, the associated $\chi^2$ values do not reliably reflect the degree of spatial structure.  The insensitivity of $\chi^2$ to array quality is demonstrated again in the histogram in Figure \ref{fig:chi_squared_hist}.  Despite a clear qualitative difference between the masked and unmasked random configurations (dashed blue and dash-dot red) the distributions of $\chi^2$ are very similar.

With the spatial dependence in mind, our next step was to develop a figure of merit based on a Bessel decomposition. The residual \UV distribution ($D(r,\phi)$, difference between actual and desired) can be decomposed into Bessel modes
\begin{equation}
%D(r,\phi) = \sum_{n=1}^{\infty} \sum_{m=0}^{\infty} J_m(k_{mn} r) (A_{mn}\sin(m\phi) + B_{mn}\cos(m\phi)),
D(r,\phi) = \sum_{n=1}^{\infty} \sum_{m=0}^{\infty} J_m\left(\frac{x_{mn} r}{R}\right) (A_{mn}\sin(m\phi) + B_{mn}\cos(m\phi)),
\end{equation}
where 
%$k_{mn}=x_{mn}/R$, and 
$x_{mn}$ is the $n^{\rm th}$ zero of the $m^{\rm th}$ Bessel function.  
The amplitudes of the asymmetric Bessel coefficients ($A_{mn},\ B_{mn},\ m>0$) reflect the asymmetric spatial over and under densities in the \UV plane. We then define a figure of merit $\zeta$ as the sum of these Bessel coefficients  
\begin{equation}
\zeta \equiv \sum_{m>0,n}^{{\rm max\ } m,n} \sqrt{A_{mn}^2 + B_{mn}^2}.
\end{equation}
Smaller $\zeta$ corresponds to less spatial structure, and hence a more desirable layout. The right hand panel of Figure \ref{fig:hists} shows the $\zeta$ distributions for the same arrays, clearly separating the masked and unmasked random arrays, and in Figure \ref{fig:examples} the $\zeta$ values accurately track the quality of the array realizations.

To minimize our figure of merit, we created an active masked algorithm based on the $\zeta$ figure of merit.  For computational reasons we first place a subset of the antennas (350 of 496 for our example) using the random masked method.  Then for each remaining antenna, we first choose a weighted random radius, $r$, and many candidate azimuthal locations (angular spacing of 10~m in our example). We then select the location with the smallest $\zeta$ value, and repeat until all $N$ antennas are placed.

The result is clear in Figures \ref{fig:examples} \& \ref{fig:hists}.  Despite having the additional constraint of the exclusion areas, the active masked method produces more symmetric baseline distributions than either the random masked (expected) or random unmasked methods (unexpected).  We can see this qualitatively by comparing the baseline distributions in the three examples.  In the thousands of arrays we examined by hand we observed a very strong correlation between small $\zeta$ and spatial symmetry .  The success of our algorithm is shown statistically by the distribution of $\zeta$ values for the three methods in Figure \ref{fig:power_hist}.  

It is of interest to note that significant baseline asymmetry arises even in the unmasked random array realizations (Figure \ref{fig:examples} top row, no exclusion areas). These asymmetries are due to shot noise in the random antenna placement. Conceptually, as the last few antennas are added images of the entire array are added to the \UV plane at that distance from the center. For centrally condensed arrays this can produce lumps in the \UV plane. Alternatively, one can consider moving a single antenna on the ground which coherently changes $N-1$ baselines. Small random associations can thus make significant correlations in the baseline distribution.

The new active method based on the Bessel decomposition figure of merit produces arrays which are superior to even an unconstrained random algorithm, even in the face of significant exclusion areas.

\section{MWA Final Layout}\label{sec:final_layout}
\begin{figure}
\begin{center}
\includegraphics{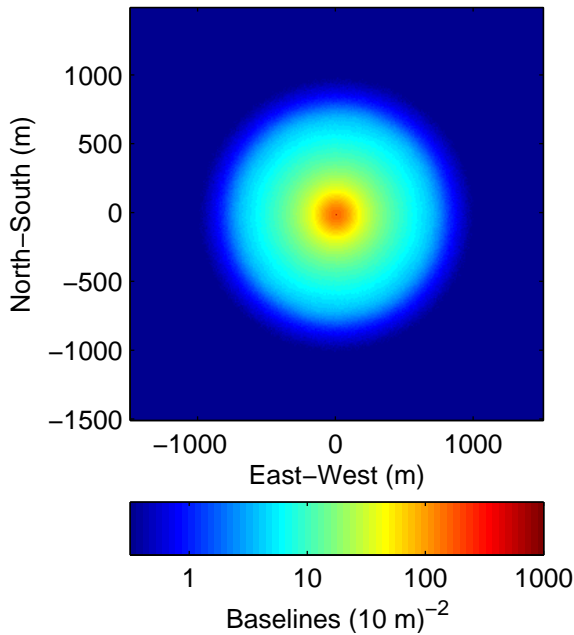}
\end{center}
\caption{MWA ideal \UV distribution.  This distribution was used in the active masked method for generating the final MWA layout.  Compare to top left pane of Figure \ref{fig:final}}
\label{fig:ideal}
\end{figure}

We have used our new algorithm to determine the final array layout of the MWA.  The full build out of the MWA will consist of 512 ``tiles'' (each comprised of 16 radio dipoles).  The majority of the tiles (496) will be distributed over a 1.5 km diameter core, with the remaining 16 tiles at a $\sim$3 km diameter to provide higher angular resolution for solar measurements.  The 16 ``outliers'' are placed by hand, while we implement our algorithm for the 496 core tiles.  The tile density distribution will be constant within a central 50m radius, and have a $r^{-2}$ dependence beyond \citep[see][for a full description of the MWA instrument]{Lonsdale:2009p3966}.  The smooth ideal distribution is shown in Figure \ref{fig:ideal}.

Several parameters of the algorithm were explored to further improve the quality of the arrays being generated.  For example, we varied the number of random tiles placed before initializing the active phase of our algorithm.  We found that placing 350 random tiles provided sufficiently unconstrained initial conditions to proceed with the active phase.  Running the algorithm in this mode 500 times provided a good sampling of the phase space.

We also investigated several array center locations within a few hundred meters of the nominal array location.  Due to the irregular distribution of avoidance areas on the ground, choosing different centers did have an effect on the quality of the best arrays generated by our algorithm.  In particular, a center near a high concentration of vegetation or rock outcrops usually results in a deficit of short baselines.  We used this information, along with feedback from a ground truth survey in February 2011 to determine our final array center.

\begin{figure}
\includegraphics{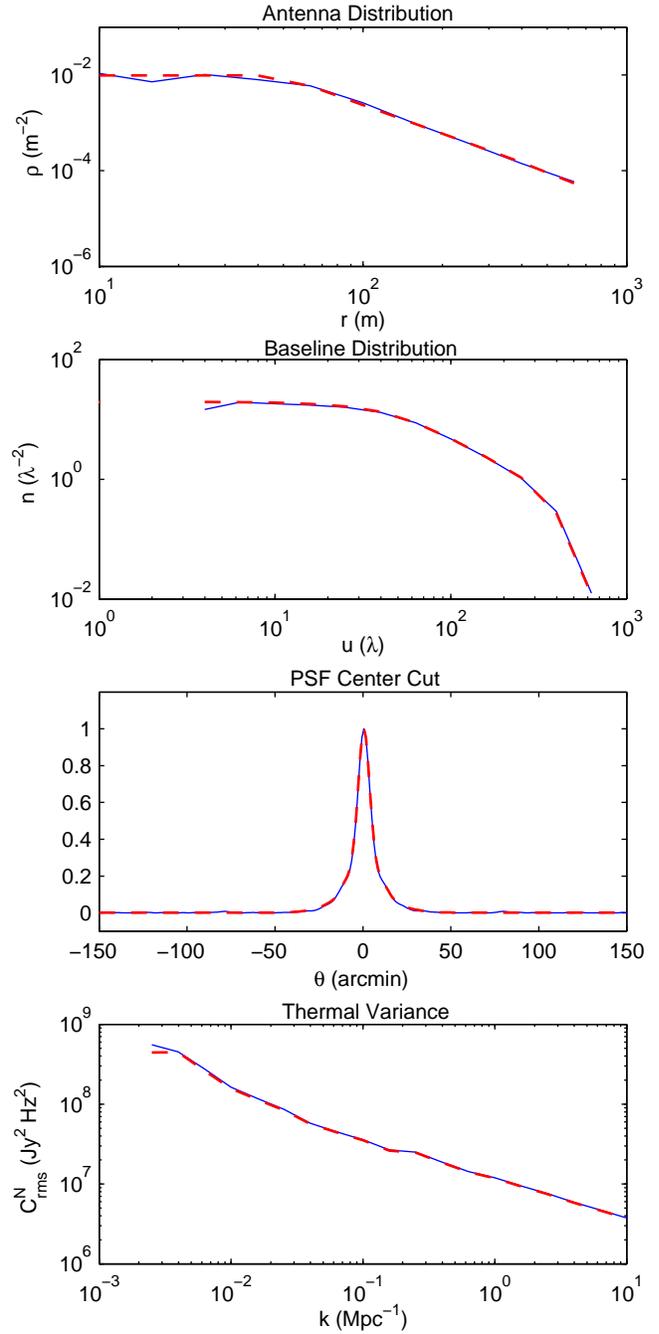}
\caption{In each figure the final MWA array layout (thin solid blue) is compared to the analytic ideal baseline distribution (thick dashed red).  Top to bottom these are the radial antenna distribution, the radial baseline distribution, a cut through the snapshot single frequency PSF, and the thermal noise as a function of cosmological wavenumber $k$ for a fiducial EoR measurement \citep[following][]{Bowman:2006p163}. In all aspects the final array very nearly traces the ideal array.}
\label{fig:stats}
\end{figure}

\begin{figure*}
\includegraphics{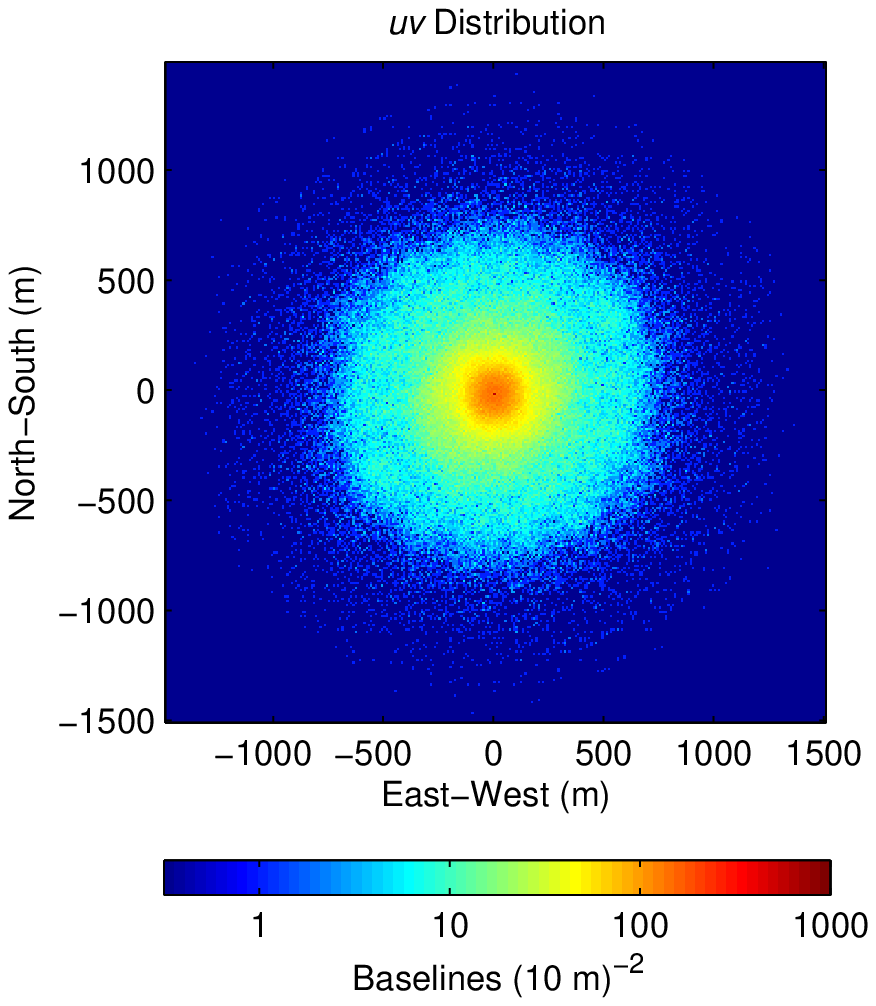}
\includegraphics{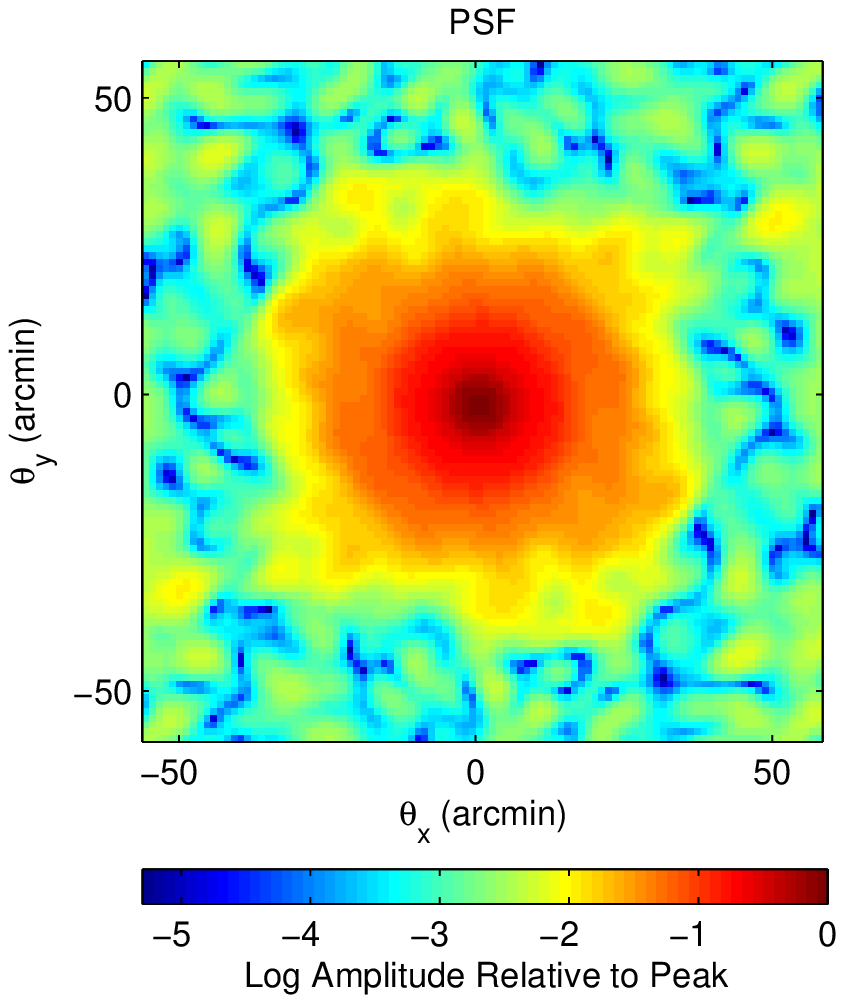}
\includegraphics{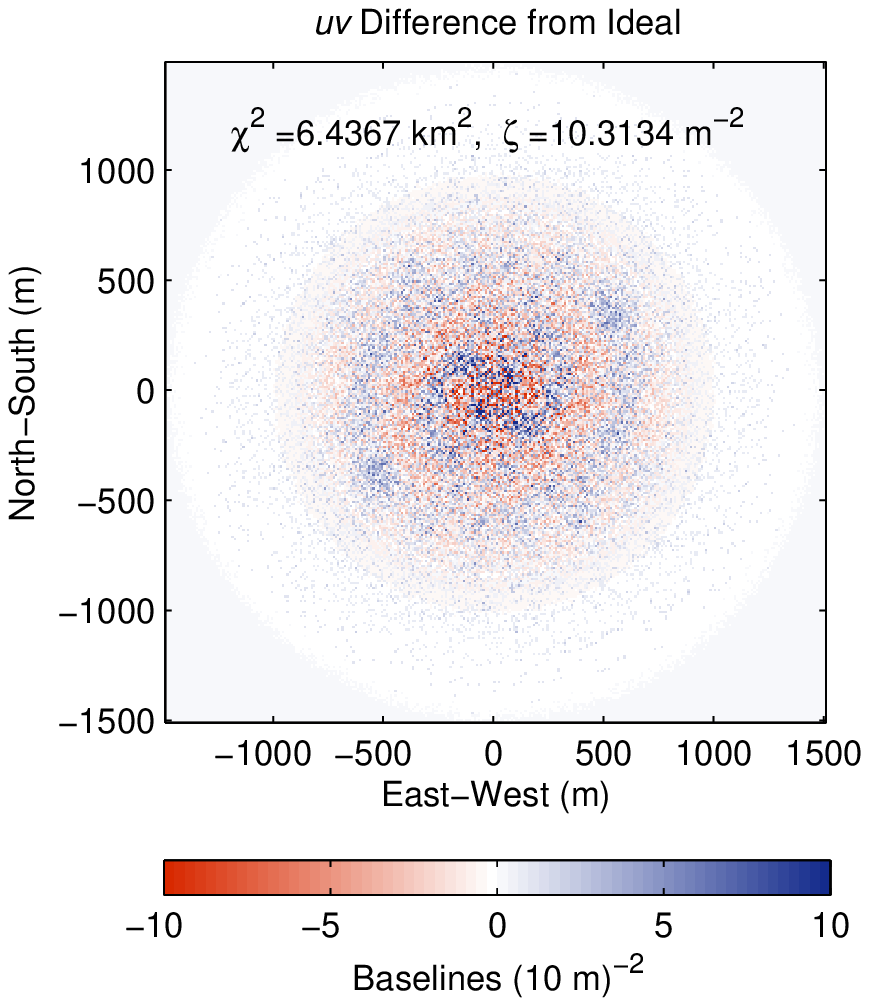}
\includegraphics{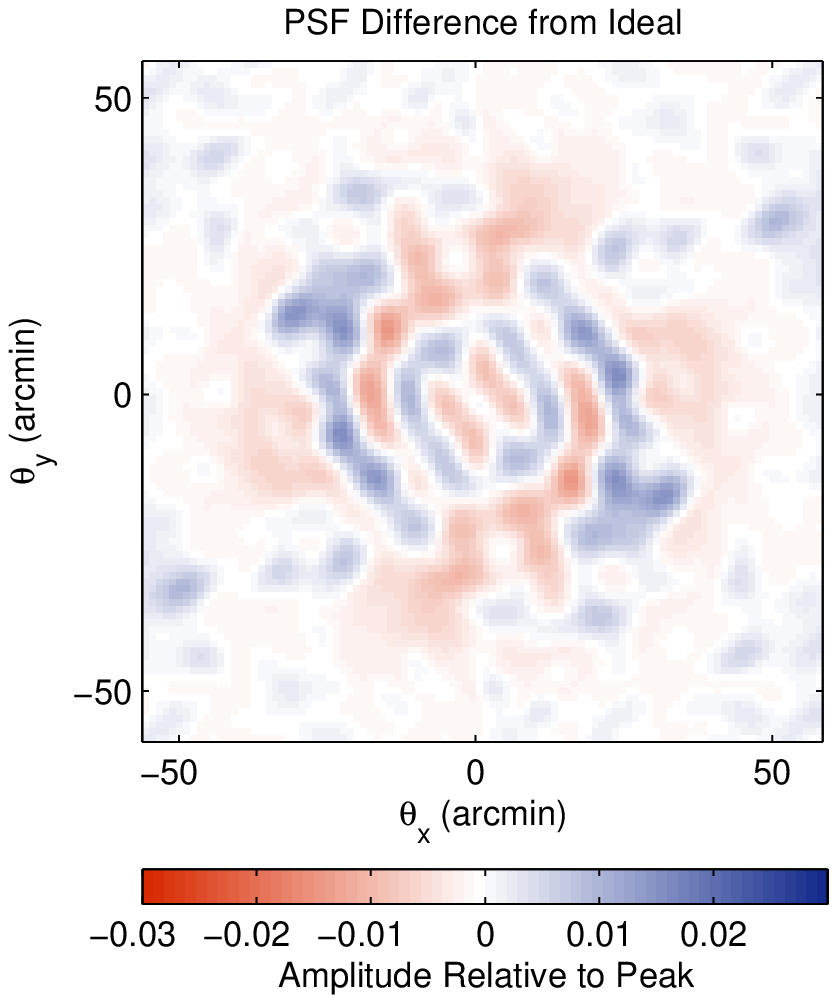}
\caption{Baseline distribution and point spread function of proposed MWA array layout.  The baseline distribution is shown in the left column, with the difference from ideal on the bottom.  The azimuthal structure is nearly completely suppressed, and only small scale noise remains.  The snapshot PSF for 150~MHz at zenith is shown in the right column.  The sensitivity relative to the peak is shown on the top, while the difference from ideal is on the bottom.  The residual ripples in the PSF difference are $\approx 1\%$ of the peak, which is on order with the background ripples expected due to our finite number of tiles.}
 \label{fig:final}
\end{figure*}

After generating 500 candidate arrays for each potential location, we used $\zeta$ as a guideline for selecting the highest quality array layouts, backed by hand grading.  The result is the final location and layout of the MWA.  Figure \ref{fig:ground_496} shows an illustration the final array overlaid on an aerial photo of the site. The locations of 496 core tiles, along with the 16 hand placed outliers, are available in the electronic supplement to this paper. Figures \ref{fig:stats} \&  \ref{fig:final} show the corresponding \UV distribution and PSF.  There is essentially no asymmetry in the final array---all of the large scale structure is greatly suppressed, providing a very smooth \UV sampling.  The small residual ripples in the PSF are on the order of the background noise we expect due to the finite number of tiles. Following the discussion of \cite{Morales:2005p796} and using the parameters from \cite{Bowman:2006p163}, the thermal noise uncertainty on the EoR power spectrum can be calculated in the bottom panel of Figure \ref{fig:stats}. With the exception of the small deviation at very low $k$, the thermal noise for our proposed layout very nearly traces that of the ideal array.

\section{Conclusions}\label{section_conclusion}
While we have been motivated by the need to generate a final array configuration for the MWA with exquisite smoothness in the PSF despite significant exclusion areas, we hope our method will be useful for determining the antenna layouts of other large $N$ arrays. In particular we have developed a new figure of merit based on Bessel decomposition that is sensitive to large scale over and under densities in the \UV plane. We have shown that algorithms based on this figure of merit can achieve extremely smooth baseline distributions while avoiding areas where antennas cannot be placed.

We have used this new algorithm to choose the final antenna locations of the MWA. The first construction stage of the MWA will consist of $\sim$128 antenna subsample of the array configuration shown in this paper.

\section*{Acknowledgments}
We acknowledge the Wajarri Yamatji people as the traditional owners of the Observatory site. We would like to particularly thank Angelica de Oliveira-Costa for helpful conversations and feedback. 

Support came from the U.S. National Science Foundation (grants AST CAREER-0847753, AST-0457585 and PHY-0835713), the Australian Research Council (grants LE0775621 and LE0882938), the U.S. Air Force Office of Scientific Research (grant FA9550-0510247), the Smithsonian Astrophysical Observatory, the MIT School of Science, the Raman Research Institute, the Australian National University, the iVEC Petabyte Data Store, the Initiative in Innovative Computing and NVIDIA sponsored Center for Excellence at Harvard, and the International Centre for Radio Astronomy Research, a Joint Venture of Curtin University of Technology and The University of Western Australia, funded by the Western Australian State government.

\bibliographystyle{apj}
\bibliography{morales}

\begin{thebibliography}{}

\bibitem[\protect\citeauthoryear{Bernardi et~al.}{Bernardi
  et~al.}{2011}]{Bernardi:2011}
Bernardi, G., Mitchell, D.~A., Ord, S.~M., Greenhill, L.~J., Pindor, B., Wayth,
  R.~B.,  \& Wyithe, J. S.~B. 2011, Monthly Notices of the Royal Astronomical
  Society, 413, 411

\bibitem[\protect\citeauthoryear{Boone}{Boone}{2001}]{Boone:2001p4841}
Boone, F. 2001, Astronomy and Astrophysics, 377, 368

\bibitem[\protect\citeauthoryear{Bowman, Morales, \& Hewitt}{Bowman
  et~al.}{2006}]{Bowman:2006p163}
Bowman, J.~D., Morales, M.~F.,  \& Hewitt, J.~N. 2006, The Astrophysical
  Journal, 638, 20 (c) 2006: The American Astronomical Society

\bibitem[\protect\citeauthoryear{Cohanim, Hewitt, \& Weck}{Cohanim
  et~al.}{2010}]{Cohanim:2010p4842}
Cohanim, B., Hewitt, J.,  \& Weck, O.~D. 2010, The Astrophysical Journal
  Supplement Series, 154, 705

\bibitem[\protect\citeauthoryear{{Cornwell}}{{Cornwell}}{1988}]{Cornwell:1988p%
1165}
{Cornwell}, T.~J. 1988, IEEE Transactions on Antennas and Propagation, 36, 1165

\bibitem[\protect\citeauthoryear{Datta, Bowman, \& Carilli}{Datta
  et~al.}{2011}]{Datta:2011p4788}
Datta, A., Bowman, J.,  \& Carilli, C. 2011, The Astrophysical Journal, 724,
  526

\bibitem[\protect\citeauthoryear{Jackson}{Jackson}{1999}]{jackson1999classical}
Jackson, J. 1999, Classical electrodynamics (Wiley)

\bibitem[\protect\citeauthoryear{Kogan}{Kogan}{2000}]{Kogan:2000p4790}
Kogan, L. 2000, IEEE Transactions on Antennas and Propagation, 48, 1075

\bibitem[\protect\citeauthoryear{Liu \& Tegmark}{Liu \&
  Tegmark}{2011}]{Liu:2011p4789}
Liu, A.,  \& Tegmark, M. 2011, Physical Review D, 83, 103006

\bibitem[\protect\citeauthoryear{Lonsdale et~al.}{Lonsdale
  et~al.}{2009}]{Lonsdale:2009p3966}
Lonsdale, C., et~al. 2009, Proceedings of the IEEE, 97, 1497

\bibitem[\protect\citeauthoryear{Morales \& Wyithe}{Morales \&
  Wyithe}{2010}]{Morales:2010p4786}
Morales, M.,  \& Wyithe, J. 2010, Annual Review of Astronomy and Astrophysics,
  48, 127

\bibitem[\protect\citeauthoryear{Morales}{Morales}{2005}]{Morales:2005p796}
Morales, M.~F. 2005, The Astrophysical Journal, 619, 678 (c) 2005: The American
  Astronomical Society

\bibitem[\protect\citeauthoryear{Morales, Bowman, \& Hewitt}{Morales
  et~al.}{2006}]{Morales:2006p147}
Morales, M.~F., Bowman, J.~D.,  \& Hewitt, J.~N. 2006, The Astrophysical
  Journal, 648, 767 (c) 2006: The American Astronomical Society

\end{thebibliography}
%\bibliography{suppbib}

%\begin{thebibliography}{99}
%\bibitem[\protect\citeauthoryear{Bowman et al.}{2006}]{Bowman_2006}Bowman J.D., Morales M.F., Hewitt J.N., 2006, ApJ, 638, 20

%\bibitem[\protect\citeauthoryear{Cohanim et al.}{2004}]{Cohanim_2004}Cohanim B.E., Hewitt J.N., de Weck O., 2004, ApJS, 154, 705

%\bibitem[\protect\citeauthoryear{Lonsdale et al.}{2009}]{Lonsdale_2009}Lonsdale C.J., Cappallo R, Morales M, Briggs F, Benkevitch L, et al., 2009, Proceedings of the IEEE 97, 1497

%\bibitem[\protect\citeauthoryear{Morales and Hewitt}{2004}]{Morales_2004}Morales M.F., Hewitt J.N., 2004, ApJ, 615, 7

%\bibitem[\protect\citeauthoryear{Morales}{2005}]{Morales_2005}Morales M.F., 2005, ApJ, 619, 678

%\end{thebibliography}

%\clearpage

%\begin{sidewaysfigure*}
\begin{landscape}
\begin{figure}
\begin{center}
%\centering
%\rule{0.75\textheight}{0.5\textheight}
\includegraphics[height = 6 in]{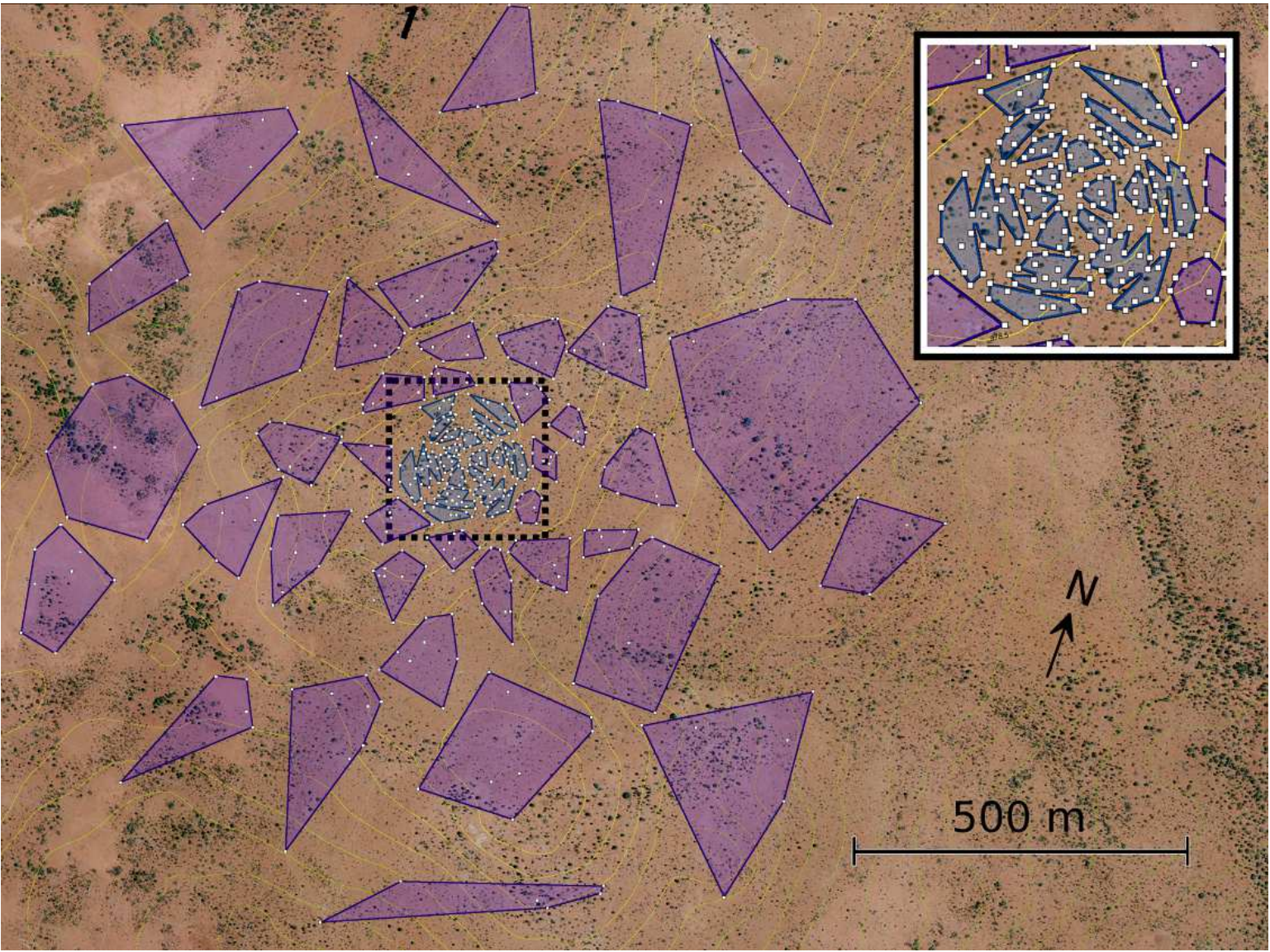}
\caption{An aerial photo of the MWA site with the final 496 tile core array superimposed.  White squares represent tiles (to scale).  The colored polygons depict a possible receiver scheme.  Each polygon outlines a receiver set's electrical footprint (8 tiles per polygon except for a few outer receiver sets that will service some of the 16 outlier tiles not shown here).  Inset: An enlarged view of the center of the array.}
\label{fig:ground_496}
\end{center}
\end{figure}
\label{lastpage}
\end{landscape}
%\end{sidewaysfigure*}
%\clearpage

\onecolumn

\begin{center}
\begin{longtable}{crr|crr|crr}
\caption{List of tile locations for 512 tile MWA layout.  The locations listed are in meters East and North relative to the center of the array at -26$^{\circ}$ 42' 4.77521'' latitude, 116$^{\circ}$ 40' 11.39333'' longitude.  The first 496 tiles were placed using our algorithm described in the paper, while the final 16 tiles were placed by hand to optimize solar measurements.}
\label{tbl:tile_locs}\\

Tile Number & East (m) & North (m) & Tile Number & East (m) & North (m) & Tile Number & East (m) & North (m)\\
\hline
\endfirsthead

\multicolumn{3}{c}%
{{Table 1 -- continued from previous page}} \\
\hline 
Tile Number & East (m) & North (m) & Tile Number & East (m) & North (m) & Tile Number & East (m) & North (m) \\
\hline
\endhead

\hline
\multicolumn{9}{r}{{Continued on next page}} \\
\endfoot

\endlastfoot

0	&	58.39	&	175.52	& 54	&	102.43	&	21.22	& 108	&	162.64	&	199.62 \\
1	&	-66.83	&	-117.18	& 55	&	622.87	&	317.32& 109	&	35.21	&	48.62 \\
2	&	4.47	&	-44.18	& 56	&	-153.15	&	82.42& 110	&	12.78	&	-36.28	\\
3	&	-22.48	&	27.72	& 57	&	58.24	&	-4.48	& 111	&	-65.47	&	-159.48	\\
4	&	-139.89	&	-49.88	& 58	&	-8.72	&	77.52& 112	&	-15.83	&	156.72	\\
5	&	105.98	&	-16.38	& 59	&	46.77	&	-75.08& 113	&	119.55	&	-3.68	\\
6	&	275.71	&	-257.68	& 60	&	-375.14	&	-116.48& 114	&	91.21	&	162.32	\\
7	&	149.63	&	-188.58	& 61	&	-144.67	&	65.82 & 115	&	-9.19	&	-73.78	\\
8	&	-589.30	&	-12.68	& 62	&	-6.06	&	-127.28 & 116	&	-45.47	&	79.32	\\
9	&	-554.53	&	-344.68	& 63	&	299.67	&	-25.78& 117	&	164.78	&	103.42	\\
10	&	-235.98	&	-97.48	& 64	&	-85.48	&	203.32 & 118	&	9.21	&	43.42	\\
11	&	-202.84	&	-39.38	& 65	&	133.51	&	59.12 & 119	&	-57.29	&	-35.28	\\
12	&	8.98	&	-73.28	& 66	&	-16.48	&	-1.08 & 120	&	16.03	&	-82.58	\\
13	&	10.81	&	-453.68	& 67	&	52.52	&	-48.98 & 121	&	-71.08	&	-110.78	\\
14	&	-251.45	&	-280.98	& 68	&	-45.67	&	24.92 & 122	&	94.94	&	-47.98	\\
15	&	187.90	&	-303.28	& 69	&	-374.81	&	149.52 & 123	&	-41.68	&	-54.48	\\
16	&	-73.07	&	-62.18	& 70	&	-536.54	&	-393.88 & 124	&	92.05	&	11.82	\\
17	&	53.00	&	-12.58	& 71	&	-100.44	&	-33.78 & 125	&	119.60	&	-58.08	\\
18	&	35.23	&	40.42	& 72	&	-67.48	&	-85.88 & 126	&	-72.90	&	-8.68	\\
19	&	304.98	&	38.42	& 73	&	-102.84	&	-16.58 & 127	&	102.10	&	-38.68	\\
20	&	11.82	&	579.82	& 74	&	-251.57	&	199.12 & 128	&	-480.41	&	190.32	\\
21	&	10.79	&	-61.58	& 75	&	-62.64	&	329.32 & 129	&	38.20	&	-33.48	\\
22	&	127.63	&	50.52	& 76	&	-308.45	&	-197.68 & 130	&	133.19	&	-625.88	\\
23	&	31.51	&	-20.58	& 77	&	-137.03	&	210.92 & 131	&	43.30	&	-24.78	\\
24	&	-207.94	&	205.52	& 78	&	8.14	&	-98.08 & 132	&	-6.34	&	-49.08	\\
25	&	-56.64	&	-60.78	& 79	&	-119.61	&	369.12 & 133	&	153.68	&	99.92	\\
26	&	486.42	&	29.52	& 80	&	24.37	&	55.52 & 134	&	-11.34	&	193.62	\\
27	&	84.02	&	3.02	& 81	&	-92.41	&	73.82 & 135	&	-25.52	&	-13.88	\\
28	&	79.49	&	-20.38	& 82	&	-84.08	&	107.32 & 136	&	-219.04	&	377.02	\\
29	&	12.74	&	-23.78	& 83	&	-5.86	&	-338.28 & 137	&	-557.56	&	-82.48	\\
30	&	-251.19	&	-107.18	& 84	&	-393.68	&	-59.68 & 138	&	-437.25	&	-64.28	\\
31	&	-18.84	&	-17.38	& 85	&	-313.56	&	151.22 & 139	&	-29.22	&	52.92	\\
32	&	634.18	&	-316.28	& 86	&	-13.62	&	28.12 & 140	&	228.82	&	195.52	\\
33	&	-471.65	&	-481.98	& 87	&	-35.07	&	17.82 & 141	&	80.32	&	50.72	\\
34	&	125.53	&	133.62	& 88	&	-163.21	&	667.92 & 142	&	-81.27	&	119.62	\\
35	&	0.58	&	17.52	& 89	&	102.92	&	-27.98 & 143	&	29.77	&	134.92	\\
36	&	-9.60	&	11.42	& 90	&	-32.97	&	127.02 & 144	&	-489.19	&	-325.98	\\
37	&	-42.31	&	-3.08	& 91	&	406.40	&	-257.48 & 145	&	72.61	&	40.52	\\
38	&	-40.26	&	-86.18	& 92	&	71.57	&	-0.78 & 146	&	-313.32	&	-630.98	\\
39	&	-48.87	&	9.02	& 93	&	141.42	&	-50.78 & 147	&	-22.83	&	2.92	\\
40	&	-69.61	&	76.42	& 94	&	99.90	&	-282.68 & 148	&	-26.87	&	-79.58	\\
41	&	-9.85	&	-6.18	& 95	&	-490.87	&	-357.48 & 149	&	-32.06	&	-41.28	\\
42	&	148.65	&	18.52	& 96	&	63.70	&	-52.68 & 150	&	-79.71	&	557.82	\\
43	&	1.20	&	-0.68	& 97	&	-0.24	&	-54.48 & 151	&	-287.67	&	-44.38	\\
44	&	-20.88	&	11.32	& 98	&	-262.10	&	356.82 & 152	&	297.58	&	49.62	\\
45	&	354.74	&	-588.58	& 99	&	38.24	&	-14.08 & 153	&	-21.16	&	-195.28	\\
46	&	61.22	&	-35.98	& 100	&	17.18	&	-1.68 & 154	&	56.35	&	104.72	\\
47	&	-648.61	&	298.82	& 101	&	74.05	&	-146.08 & 155	&	-281.52	&	-163.38	\\
48	&	-66.65	&	-140.18	& 102	&	234.14	&	137.32 & 156	&	30.55	&	-37.88	\\
49	&	28.34	&	-7.58	& 103	&	47.76	&	40.02 & 157	&	96.44	&	130.42	\\
50	&	103.83	&	-106.78	& 104	&	-165.60	&	27.52 & 158	&	-40.07	&	-17.68	\\
51	&	90.90	&	220.02	& 105	&	150.53	&	86.22 & 159	&	43.78	&	-143.08	\\
52	&	-357.63	&	-234.38	& 106	&	-81.92	&	-78.18 & 160	&	17.89	&	-52.48	\\
53	&	-52.88	&	-3.18	& 107	&	-493.65	&	-160.08 & 161	&	25.44	&	70.72	\\
162	&	78.12	&	13.42	&	221	&	49.46	&	49.42	&	280	&	-5.95	&	-17.38	\\		
163	&	106.61	&	27.82	&	222	&	-80.73	&	66.32	&	281	&	23.6	&	23.92	\\		
164	&	47.4	&	-61.08	&	223	&	1.26	&	-9.48	&	282	&	-24.26	&	-90.88	\\		
165	&	289.06	&	-6.68	&	224	&	-14.15	&	-26.18	&	283	&	50.15	&	-31.78	\\		
166	&	35.91	&	-70.48	&	225	&	424.95	&	-16.68	&	284	&	-240.33	&	-115.28	\\		
167	&	4.9	&	57.82	&	226	&	116.13	&	-105.58	&	285	&	-144.03	&	-222.98	\\		
168	&	-9.82	&	37.32	&	227	&	582.58	&	151.22	&	286	&	-614.08	&	57.72	\\		
169	&	10.85	&	95.42	&	228	&	25.22	&	36.12	&	287	&	-155.48	&	507.72	\\		
170	&	-5.71	&	-105.28	&	229	&	59.36	&	40.02	&	288	&	48.45	&	-41.78	\\		
171	&	87.85	&	-25.48	&	230	&	-32.56	&	33.22	&	289	&	16.46	&	73.92	\\		
172	&	-221.52	&	-223.08	&	231	&	210.48	&	-439.08	&	290	&	89.41	&	-108.38	\\		
173	&	-48.19	&	-78.58	&	232	&	-143.52	&	126.82	&	291	&	599.01	&	-482.78	\\		
174	&	-56.86	&	22.62	&	233	&	121.07	&	-85.18	&	292	&	-13.43	&	-36.98	\\		
175	&	110.39	&	-486.58	&	234	&	-462.11	&	228.62	&	293	&	240.24	&	-36.08	\\		
176	&	471.84	&	-257.48	&	235	&	-14.5	&	113.52	&	294	&	179.43	&	296.12	\\		
177	&	99.39	&	46.42	&	236	&	-425.82	&	-347.78	&	295	&	245.99	&	-412.58	\\		
178	&	135.91	&	-12.08	&	237	&	-11.04	&	-61.68	&	296	&	15.83	&	14.82	\\		
179	&	40.07	&	-98.38	&	238	&	31.64	&	100.32	&	297	&	-513.27	&	-51.18	\\		
180	&	-46.48	&	-203.18	&	239	&	56.33	&	62.02	&	298	&	-399.94	&	126.92	\\		
181	&	-126.63	&	-127.48	&	240	&	27.69	&	12.82	&	299	&	269.85	&	-17.68	\\		
182	&	581.84	&	201.12	&	241	&	382.27	&	395.72	&	300	&	65.11	&	32.22	\\		
183	&	128.74	&	40.22	&	242	&	-525.74	&	-283.68	&	301	&	-196.28	&	-113.18	\\		
184	&	260.35	&	-481.48	&	243	&	-590.07	&	89.02	&	302	&	333.53	&	-17.68	\\		
185	&	-40.22	&	10.72	&	244	&	-258.51	&	-190.68	&	303	&	240.92	&	33.82	\\		
186	&	312.54	&	-10.78	&	245	&	85.09	&	-44.28	&	304	&	105.76	&	-191.88	\\		
187	&	48.58	&	30.82	&	246	&	-531.15	&	-470.58	&	305	&	225.21	&	-74.68	\\		
188	&	331.12	&	46.22	&	247	&	-476.19	&	122.92	&	306	&	79.97	&	-54.98	\\		
189	&	60.6	&	82.62	&	248	&	78.67	&	141.22	&	307	&	-50.76	&	178.52	\\		
190	&	-395.33	&	380.02	&	249	&	172.95	&	375.42	&	308	&	-266.16	&	-38.58	\\		
191	&	234.43	&	279.22	&	250	&	198.46	&	-131.88	&	309	&	-189.04	&	-301.88	\\		
192	&	-89.52	&	-53.98	&	251	&	-38.16	&	73.92	&	310	&	65.74	&	72.02	\\		
193	&	40.78	&	74.42	&	252	&	-0.09	&	46.82	&	311	&	23.73	&	-134.98	\\		
194	&	-299.89	&	145.02	&	253	&	-34.85	&	-157.58	&	312	&	-80.92	&	-38.88	\\		
195	&	68.31	&	-221.98	&	254	&	3.3	&	-66.08	&	313	&	27.27	&	47.82	\\		
196	&	13.54	&	6.02	&	255	&	39.7	&	588.02	&	314	&	58.93	&	246.82	\\		
197	&	-337.13	&	-194.58	&	256	&	-89.29	&	-65.18	&	315	&	-29.18	&	-2.38	\\		
198	&	135.05	&	-144.38	&	257	&	-530.37	&	-297.28	&	316	&	-375.11	&	-37.38	\\		
199	&	-10.56	&	-264.98	&	258	&	17.42	&	59.52	&	317	&	-102.4	&	-128.38	\\		
200	&	150.66	&	-284.88	&	259	&	208.81	&	64.02	&	318	&	42.57	&	-7.08	\\		
201	&	-141.84	&	203.02	&	260	&	-15.35	&	-68.28	&	319	&	-12.78	&	46.22	\\		
202	&	-41.72	&	-25.58	&	261	&	68.05	&	13.62	&	320	&	-53.03	&	-49.08	\\		
203	&	106.17	&	213.22	&	262	&	-42.97	&	152.02	&	321	&	262.5	&	134.62	\\		
204	&	-105.04	&	-136.38	&	263	&	67.19	&	24.12	&	322	&	60.09	&	-23.48	\\		
205	&	212.32	&	34.92	&	264	&	48.64	&	64.22	&	323	&	18.28	&	49.32	\\		
206	&	-228.13	&	-172.18	&	265	&	131.22	&	293.52	&	324	&	-120.84	&	186.22	\\		
207	&	188.44	&	-100.18	&	266	&	40.26	&	63.32	&	325	&	-291.93	&	-267.98	\\		
208	&	-588.48	&	-201.09	&	267	&	-286.24	&	401.72	&	326	&	619.19	&	-146.18	\\		
209	&	-536.5	&	186.72	&	268	&	-136.15	&	159.82	&	327	&	-445.37	&	305.82	\\		
210	&	-53.13	&	202.02	&	269	&	28.93	&	-123.18	&	328	&	-6.98	&	20.72	\\		
211	&	158.01	&	86.82	&	270	&	-21.67	&	97.42	&	329	&	-452.88	&	378.52	\\		
212	&	477.21	&	429.92	&	271	&	-49.84	&	-192.28	&	330	&	61.33	&	127.82	\\		
213	&	-17.61	&	-43.78	&	272	&	91.37	&	24.82	&	331	&	-133.43	&	674.12	\\		
214	&	-26.72	&	-24.08	&	273	&	-2.21	&	67.62	&	332	&	26.2	&	1.92	\\		
215	&	8.68	&	-11.18	&	274	&	-232.93	&	-431.88	&	333	&	46.65	&	88.52	\\		
216	&	-400.7	&	-276.08	&	275	&	30.49	&	-107.18	&	334	&	152.1	&	467.72	\\		
217	&	-300.77	&	-67.28	&	276	&	70.12	&	-36.68	&	335	&	-27.94	&	8.72	\\		
218	&	-96.96	&	358.02	&	277	&	44.92	&	0.52	&	336	&	92.67	&	69.72	\\		
219	&	60.48	&	9.52	&	278	&	183.34	&	351.22	&	337	&	168.61	&	81.72	\\		
220	&	335.63	&	-582.78	&	279	&	288.72	&	-36.38	&	338	&	144.65	&	313.02	\\		
339	&	95.55	&	207.82	&	397	&	3.23	&	96.32	&	455	&	45.09	&	107.12	\\		
340	&	142.88	&	-191.88	&	398	&	-73.6	&	130.42	&	456	&	-65.92	&	349.52	\\		
341	&	-100.95	&	539.02	&	399	&	-28.39	&	-54.58	&	457	&	142.17	&	131.92	\\		
342	&	32.81	&	5.72	&	400	&	172.36	&	-130.88	&	458	&	-54.46	&	42.32	\\		
343	&	607.76	&	-227.98	&	401	&	33.23	&	26.72	&	459	&	88.04	&	-14.08	\\		
344	&	47.11	&	-18.38	&	402	&	85.25	&	42.42	&	460	&	11.78	&	54.22	\\		
345	&	144.37	&	237.02	&	403	&	7.33	&	-168.38	&	461	&	-424.49	&	408.52	\\		
346	&	169.95	&	-233.48	&	404	&	-82.98	&	53.32	&	462	&	20.1	&	-38.48	\\		
347	&	71	&	81.22	&	405	&	15.86	&	169.12	&	463	&	258.06	&	-50.48	\\		
348	&	-151.05	&	34.72	&	406	&	719.42	&	159.62	&	464	&	66.1	&	249.22	\\		
349	&	325.26	&	134.32	&	407	&	40.35	&	-725.78	&	465	&	233.9	&	-27.28	\\		
350	&	-178.36	&	117.52	&	408	&	-64.67	&	-201.38	&	466	&	-183.86	&	-215.58	\\		
351	&	-48.57	&	-32.68	&	409	&	-160.11	&	16.12	&	467	&	0.09	&	187.02	\\		
352	&	-54.42	&	-10.68	&	410	&	16.61	&	35.12	&	468	&	347.9	&	366.92	\\		
353	&	-152.44	&	108.92	&	411	&	-231.14	&	113.42	&	469	&	-44.8	&	-643.38	\\		
354	&	-70.55	&	94.62	&	412	&	119.22	&	41.22	&	470	&	18.62	&	-64.08	\\		
355	&	11.56	&	-303.88	&	413	&	33.48	&	-240.48	&	471	&	-95.84	&	144.92	\\		
356	&	-21.69	&	40.92	&	414	&	-131.83	&	-25.88	&	472	&	199.12	&	246.92	\\		
357	&	19.55	&	-27.48	&	415	&	52.56	&	22.32	&	473	&	55.01	&	14.82	\\		
358	&	8.46	&	24.02	&	416	&	43.12	&	17.72	&	474	&	-118.97	&	-410.78	\\		
359	&	-92.55	&	98.42	&	417	&	318.39	&	-296.28	&	475	&	-35.21	&	-8.78	\\		
360	&	-37.58	&	61.62	&	418	&	-53.98	&	-205.78	&	476	&	649.05	&	20.32	\\		
361	&	-44.58	&	52.02	&	419	&	238.65	&	579.92	&	477	&	63.56	&	-65.78	\\		
362	&	-1.92	&	-479.48	&	420	&	80.74	&	-28.78	&	478	&	92.22	&	-5.58	\\		
363	&	91.63	&	2.72	&	421	&	-54.89	&	62.02	&	479	&	135.72	&	726.32	\\		
364	&	-2.72	&	-396.08	&	422	&	-21.69	&	-352.58	&	480	&	-31.59	&	24.92	\\		
365	&	219.91	&	-132.58	&	423	&	-271.9	&	43.72	&	481	&	-14.55	&	19.62	\\		
366	&	33.38	&	-63.18	&	424	&	270.06	&	287.42	&	482	&	213.08	&	-35.68	\\		
367	&	-270.64	&	170.12	&	425	&	42.1	&	53.02	&	483	&	73.65	&	-114.18	\\		
368	&	-29.34	&	112.62	&	426	&	37.6	&	-42.58	&	484	&	394.22	&	-282.08	\\		
369	&	265.52	&	-121.48	&	427	&	-53.96	&	33.42	&	485	&	-12.49	&	-80.68	\\		
370	&	-3.14	&	6.82	&	428	&	-357.63	&	488.02	&	486	&	48.35	&	-160.18	\\		
371	&	-119.71	&	142.42	&	429	&	-9.83	&	2.42	&	487	&	-34.38	&	-74.78	\\		
372	&	-143.07	&	150.72	&	430	&	-178.78	&	-28.28	&	488	&	-162.27	&	-148.58	\\		
373	&	76.09	&	23.02	&	431	&	43.78	&	-276.18	&	489	&	75.89	&	-63.68	\\		
374	&	-53.09	&	75.12	&	432	&	-36.44	&	50.52	&	490	&	65.88	&	-44.08	\\		
375	&	16.81	&	-12.38	&	433	&	-223.65	&	64.02	&	491	&	-491.74	&	107.62	\\		
376	&	407.46	&	522.12	&	434	&	56.92	&	32.72	&	492	&	129.98	&	-29.28	\\		
377	&	-74.12	&	110.92	&	435	&	66.25	&	146.22	&	493	&	-206.43	&	480.22	\\		
378	&	42.34	&	-50.58	&	436	&	-28.41	&	61.42	&	494	&	-187.87	&	-419.78	\\		
379	&	422.28	&	-540.48	&	437	&	-52.12	&	313.52	&	495	&	102.43	&	5.22	\\		
380	&	-45.53	&	-11.68	&	438	&	66.11	&	53.62	&	496	&	-241.24	&	1432.02	\\		
381	&	335.51	&	527.72	&	439	&	-165.37	&	166.52	&	497	&	97.41	&	1187.42	\\		
382	&	104.51	&	592.02	&	440	&	327.84	&	-315.38	&	498	&	808.32	&	1183.41	\\		
383	&	666.25	&	137.52	&	441	&	-182.09	&	-278.78	&	499	&	932.9	&	768.93	\\		
384	&	-110.83	&	136.32	&	442	&	71.8	&	-21.58	&	500	&	1405.56	&	336.61	\\		
385	&	422.64	&	-528.28	&	443	&	108.3	&	95.12	&	501	&	1197.9	&	-67.42	\\		
386	&	13.73	&	-382.38	&	444	&	-154.65	&	-484.48	&	502	&	1296.16	&	-662.28	\\		
387	&	-171.85	&	-466.18	&	445	&	-283.9	&	411.92	&	503	&	545.29	&	-1053.82	\\	
388	&	-4.59	&	-25.58	&	446	&	-60.57	&	-396.28	&	504	&	749.97	&	-1255.22	\\		
389	&	33.14	&	-310.08	&	447	&	104.72	&	-360.28	&	505	&	-380.8	&	-1036.16	\\		
390	&	-3.02	&	30.92	&	448	&	-59.93	&	6.12	&	506	&	-477.74	&	-1347.34	\\		
391	&	-38.45	&	90.92	&	449	&	196.65	&	-54.18	&	507	&	-584.46	&	-854.81	\\		
392	&	200.5	&	-114.58	&	450	&	-521.4	&	259.62	&	508	&	1268.36	&	641.4	\\		
393	&	24.91	&	-59.28	&	451	&	333.52	&	-171.88	&	509	&	-983.33	&	-138.74	\\		
394	&	327.49	&	546.82	&	452	&	-176.63	&	-37.18	&	510	&	-926.52	&	354.91	\\		
395	&	-5.64	&	-40.58	&	453	&	326.58	&	596.92	&	511	&	-795.88	&	862.29	\\		
396	&	155.24	&	592.02	&	454	&	24.07	&	-79.78	\\								
\hline
\end{longtable}
\end{center}

\end{document}